\documentclass[a4paper,11pt]{article}
\usepackage{graphicx} 
\usepackage{authblk}
\usepackage[dvipsnames]{xcolor}

\usepackage[hypertexnames=false]{hyperref}
\hypersetup{
    colorlinks=true,
    linkcolor=blue,
    filecolor=magenta,   
    citecolor=black,   
    urlcolor=cyan,
    }

\def\blu#1 {{\color{ForestGreen}{Jarah: #1}}}

\title{Testing the Gallium anomaly using Electron-Neutrino Scattering}

\author[a]{Emilio Ciuffoli\footnote{emilio@impcas.ac.cn}}
\author[a,b]{Jarah Evslin\footnote{jarah@impcas.ac.cn}}
\author[c]{Ruixuan Gao}
\author[c]{Haixing Lin}
\author[c]{Jian Tang\footnote{Corresponding author: tangjian5@mail.sysu.edu.cn }}

\affil[a]{Institute of Modern Physics, NanChangLu 509, Lanzhou 730000, China}
\affil[b]{University of the Chinese Academy of Sciences, YuQuanLu 19A, Beijing 100049, China}
\affil[c]{School of Physics, Sun Yat-sen University, Guangzhou 510275, China}

\begin{document}

\maketitle

\begin{abstract}

\noindent
The Gallium anomaly is an unexplained deficit in the neutrinos 
observed during the calibration of GALLEX and SAGE using a $^{51}$Cr radioactive source and recently confirmed by BEST.  The possible explanations for this deficit include an overestimation of the neutrino absorption cross section in Ga, an incorrect measurement of the source activity or the existence of sterile neutrinos. However, as this deficit has only been observed in Ga detectors, it has not been possible to distinguish among various proposals. Therefore, we propose an experiment using the same radioactive source but with a different detection method, electron-neutrino scattering. 
We discuss potential locations for such an experiment, estimating the main backgrounds and expected event rates, considering various target masses and source positions. Even if the anomaly does not result from the detection method, such an experiment can provide an independent determination of the branching ratio of the $^{51}$Cr decay by using the spectral information or observing the scattering angle.  
It is also sensitive to an eventual baseline dependence of the anomaly, as is predicted in sterile neutrino models.

\end{abstract}

\section{Introduction}

\setcounter{footnote}{0}

The GALLEX \cite{GALLEX:1992gcp} and SAGE \cite{SAGE:1994ctc} experiments began operations in the late 1980s, continuing up to the beginning of the following century.  Their goal was to study solar neutrinos using a Gallium-based detector, observing $\nu_e$ via the reaction
\begin{equation}\label{Eq::ReactionGallium}
    ^{71}\mathrm{Ga}+\nu_e \rightarrow \textrm{ }^{71}\mathrm{Ge} + e^-.
\end{equation}
The main advantage of such a method is the very low energy threshold of (\ref{Eq::ReactionGallium}), namely 234 keV, which allowed the experiments to study neutrinos from the low-energy pp chain.

Radioactive sources were used to calibrate the detectors,  most notably $^{51}$Cr\footnote{For SAGE, $^{37}$Ar was used as well.}, which primarily decays via electron capture
\begin{equation}\label{Eq::CrDecay}
    ^{51}\mathrm{Cr}\rightarrow \textrm{ }^{51}\mathrm{V} + \nu_e.
\end{equation}
The neutrino emitted is monoenergetic: in $\sim 90\%$ if the cases  its energy is $E_{\nu}=E_{\nu,2}\sim$750 keV.  In about $10\%$ of the cases, we have $E_{\nu}=E_{\nu,1}\sim$430 keV and a 320 keV $\gamma$ is emitted as well
\begin{equation}\label{Eq::CrDecayGamma}
    ^{51}\mathrm{Cr}\rightarrow \textrm{ }^{51}\mathrm{V} + \nu_e+\gamma.
\end{equation}
During the calibration of the GALLEX and SAGE detectors, only about $80\%$ of the expected neutrino flux was observed. More recently, in 2022, this anomalous deficit was confirmed by the BEST experiment \cite{Barinov:2021asz} (see also Ref. \cite{Elliott:2023cvh} for a comprehensive review of the experimental results on the Gallium anomaly).  Currently the measured deficit, expressed as the ratio between the observed rate and the expected one, is $R=N_{obs}/N_{exp}=0.803 \pm 0.035$ \cite{Brdar:2023cms}.

Many explanations for this deficit have been proposed. However, they are all in tension with other experimental results, require new physics, or both (see \cite{Brdar:2023cms} for a review).  We can roughly divide these proposed explanations into three categories: those involving the neutrino production, detection or propagation.

The source of the anomaly could, for example, be an error in the estimation of the source activity. The measurements are performed by detecting the $\gamma$'s emitted in $\sim$10\% of the cases: an error in the branching ratio (BR) for those decays would lead to an overestimation of the source activity, explaining the anomaly. On the other hand, these branching ratios have already been measured by several collaborations \cite{Fisher:1984rrq,Cr-2,Cr-3,Cr-4}, with precisions of the order of $10^{-4}$.

Another possible class of explanations involves some errors in the estimation of the detection rate and/or efficiency.  For example, the cross-section for the reaction (\ref{Eq::ReactionGallium}) may be underestimated, or there may be an error in the radiochemical extraction of $^{71}$Ge from the detector, which is used to estimate the total number of neutrinos detected. Recently, however, Ref. \cite{Gonzalez-Garcia:2023kva} has compared the solar neutrino fluxes measured by the Gallium experiments with the ones measured by Borexino, finding a substantial agreement between the two datasets.  If the explanation of the anomaly relies on the detection method, one would expect that the same deficit should apply to the measurement of the solar neutrinos as well, but this does not appear to be the case. 

A third class of explanations involves the propagation process, the most famous of which is sterile neutrinos. This explanation gained some popularity a few years ago when it seemed possible to simultaneously explain several neutrino anomalies (although a "pragmatic" approach would be needed \cite{Giunti:2013aea}, since the sterile neutrinos would not be able to explain the low-energy excess seen at MiniBooNE) with just a single flavor of sterile neutrinos. 

The hopes for a simple explanation, however, were soon quashed. First of all, the region of the parameter space that could explain the Gallium anomaly is quite strongly excluded by short baseline experiments, which failed to see signs of sterile neutrinos \cite{Giunti:2022btk}. Moreover, it became clear recently that the reactor anomaly was due to an error in the theoretical prediction of the neutrino flux; even if the origin of the shape distortion at around 5 MeV is still unclear, it cannot be due to the presence of sterile neutrinos. Finally, in the $\nu_\mu\rightarrow \nu_e$ sector, the low-E neutrino excess cannot be explained by oscillations alone (with or without sterile neutrinos), and the search for new systematics that can justify the so-called excess has not produced any results yet. 

It should also be noted that the relevant bounds on the mixing parameters that would exclude the sterile neutrino explanation are obtained using $\bar{\nu}_e$, while the Gallium anomaly was measured in the neutrino sector. Even though it's true that a CP-violating phase could introduce a difference between the oscillation probability in the neutrino and antineutrino sector, such a difference cannot be present in the survival probability, namely the oscillation probability for $\bar{\nu}_e\rightarrow \bar{\nu}_e$ must be the same as for $\nu_e\rightarrow\nu_e$, while the probability for $\bar{\nu}_\mu\rightarrow \bar{\nu}_e$ could be different from the one for $\nu_\mu\rightarrow\nu_e$. Since in the Gallium anomaly we are observing the survival probability, {\it i.e.} the former, if sterile neutrinos are the cause of the Gallium anomaly, CP violation would not be sufficient to explain the tension with the short baseline experiments, CPT must be broken as well \cite{Giunti:2010zs}.

This is why, recently, the focus of the neutrino community shifted: instead of attempting to find a global explanation for all the anomalies, we should try first to solve "locally" the inconsistencies in each channel separately, looking for an explanation for the contradictory results \cite{Maltoni}. This, however, requires new experimental data, which is the main motivation for our work. 

We propose a test of the Gallium anomaly, still using $^{51}$Cr as a source but detecting the neutrinos via electron-neutrino scattering, rather than neutrino capture.  If the Gallium anomaly persists without Gallium, we can exclude any explanation involving the detector. This is similar to the ill-fated SOX experiment \cite{Borexino:2013xxa}, which was proposed and then abandoned due to issues with the delivery of the $^{144}$Ce radioactive source.  For this project, however, we will need only $^{51}$Cr, which has already been used successfully in several experiments, including GALLEX, SAGE and BEST. A similar setup has also been proposed to study the neutrino anapole moment \cite{Herrera:2024ysj}.

The main objective of this endeavor would be to measure the total neutrino flux.  This would be sufficient prove or exclude any explanations involving the detection process. A secondary and more challenging goal would be to directly measure the BR of the $^{51}$Cr decay, which would allow us also to test the conjecture that the anomaly is caused by the calibration of the source. The BR would be determined by looking at the energy spectrum of the recoiled electron or, if we use Cherenkov light, by reconstructing the scattering angle which, along with the recoil energy, would allow us to reconstruct the neutrino energy. The requirements that must be satisfied to carry out this measurement, however, are much steeper than the ones needed to simply confirm or deny the presence of the anomaly.  The precision that can be achieved measuring the BR using electron-neutrino scattering is considerably worse than that obtained in the previously cited works.  We feel that this is nonetheless of interest because an error in the BR is often mentioned as possible explanation of the anomaly, and this kind of measurement would provide an independent test of this hypothesis.
Finally, using a liquid scintillator it would also be possible to obtain information on the spatial distribution of the events, which would allow us to see an eventual dependence of the anomaly on the baseline.  If observed, such an effect would be a strong hint in favor of sterile neutrinos.

This work is organized as follows: in Sec.~\ref{sec::EnuScattering} we will describe how one may use electron-neutrino scattering to test Gallium anomaly.  In Sec.~\ref{sec::Experiments} we will consider possible locations where such an experiment may be performed, computing the expected number of events and the main backgrounds.  In Sec.~\ref{sec::BR} we will show how it would be possible to measure the BR of $^{51}$Cr decay using such an experiment, discussing the requirements needed to carry out such a measurement.  In Sec.~\ref{sec::Sterile} we will compute the region of the parameter space that can be tested if the anomaly is caused by sterile neutrinos. Finally, in Sec.~\ref{sec::Conclusions}, we will summarize our results.

\section{Electron-Neutrino Scattering}
\label{sec::EnuScattering}
\subsection{Cross Section}

$^{51}$Cr decays via electron capture, which means that a monoenergetic neutrino will be emitted. In the  majority of the cases ($\sim 90\%)$, the neutrino will be the only particle emitted, with energy $E_{\nu,2}=750 $ keV. In a fraction of the decays ($p_{\mathrm{BR}}\sim 10\%$), a 320 keV $\gamma$ will be emitted as well, lowering the neutrino energy to $E_{\nu,1}=430$ keV. This decay mode is extremely important because the activity of the source is measured by detecting the $\gamma$'s, which means that even a small error in the BR could significantly change the number of detections and so potentially explain the anomaly. 

The main challenge that must be overcome by any experiment that wants to use $^{51}$Cr as a neutrino source is the very low energy of the $\nu_e$, which excludes most of the methods used to detect relatively low-energy neutrinos, such as Inverse Beta Decay.

It is possible, however, to see those neutrinos using elastic electron neutrino scattering.  The cross-section for this process is \cite{Tomalak:2019ibg}
\begin{eqnarray}\label{eq::DefSigmaT}
\frac{d\sigma}{d T}(T,E_{\nu}) &=& \frac{2G_F^2 Z m_e}{\pi}\left(C_L^2+C_R^2\left(1-\frac{T}{E_{\nu}}\right)^2-C_LC_R\frac{m}{E_\nu}\frac{T}{E_\nu}\right)\theta(T_{max}-T) \nonumber \\
&& C_L=\textrm{Sin}^2(\theta_W)+\frac{1}{2} \qquad C_R=\textrm{Sin}^2(\theta_W) 
\end{eqnarray}
where $G_F$ is the Fermi constant, $m_e$ the electron mass, $\theta_W$ the weak mixing angle, $\theta(x)$ the Heaviside function, $E_\nu$ the neutrino energy and $T_{(max)}$ the (maximum) electron recoil energy. $T_{max}$ is given by
\begin{equation}
    T_{max}(E_\nu)=\frac{2E_\nu^2}{2E_\nu+m_e}.
\end{equation}
It should also be underlined that, for a generic value of $E_\nu$, the cross-section does not go to zero when $T$ approaches $T_{max}$.

If $^{51}$Cr is used as neutrino source, the maximum recoil energies would be quite low (albeit still largely within the sensitivity of modern technology)
\begin{equation}
    T_{max,1}=T_{max}(E_{\nu,1})\simeq 270 \textrm{ keV}  \qquad T_{max,2}=T_{max}(E_{\nu,2})\simeq 560 \textrm{ keV}. 
\end{equation}
For this reason, we will consider liquid scintillator detectors: they don't have, in principle, a low-energy threshold, nonetheless one should be imposed due to other factors. For example, if the background is dominant at low energies, then an effective threshold could be {\it de facto} present.  Moreover, in order to measure the scattering angle, it is necessary to observe the Cherenkov light emitted by the electron, which would not be present if its kinetic energy is lower than 200 keV\footnote{The actual threshold for the emission of Cherenkov light is a bit lower, around 176 keV, however 200 keV can be used as practical lower bound.} (this will be discussed in detail in Secs.~\ref{sec::Experiments} and \ref{sec::BR}).

\section{Detector}
\label{sec::Experiments}
It is straightforward to obtain a lower bound on the event rate needed in order to test the Gallium anomaly.  Letting $N_{tot}$ be the expected number of events (considering the efficiency of the detector, the low-energy threshold, {\it{etc.}}..,) and assuming that the background is negligible, one finds
\begin{equation}
    \sigma_{Ga}=\frac{1}{\sqrt{N_{tot}}}.
\end{equation}
As the anomaly is currently of the order of 20\%, to exclude it at 5$\sigma$ one must observe at least 625 events.

The main factor limiting $N_{tot}$ will be the running time of the experiment, which will be constrained by the half-life of the source. Indeed, since the half-life of $^{51}$Cr is $\sim$ 27.7 days, there is only a limited window of time in which data can be collected. For this reason, we will consider running times between 10 and 30 days.  Note that, due to source depletion, these would respectively be equivalent to 8.85 and 21.1 days with the source at full activity. In the Ga experiments, the running time of the calibration was considerably shorter, of the order of 5-10 days, because they were also constrained by the half-life of $^{71}$Ge, which is 11 days.  We would not have such a limitation using electron-neutrino scattering.

$^{51}$Cr is usually created from $^{50}$Cr via neutron irradiation. $^{50}$Cr is not a particularly rare isotope (its natural abundance is 4.34\%), however, since the neutron absorption cross section of $^{53}$Cr (natural abundance: 0.5\%) is comparable to that of $^{50}$Cr (18.1 b vs 15.8 b, respectively \cite{Sears}), the sample should be depleted of the former and enriched of the latter before irradiation. In BEST, a 4-kg sample of $^{50}$Cr was used, with an isotopic purity of 97\%, the activity of the source was measured to be 1.16$\times 10^{17}$ Bq (or, equivalently, 3.14 MCi). For simplicity, in this paper we will assume a source activity of $10^{17}$ Bq.

We should also point out that the expected number of events will not be proportional to the detector mass. Indeed, if we consider a point-like source and a detector of finite dimension, the number of events will be proportional to\footnote{as long as no baseline-dependent effects are present, such as sterile neutrinos}
\begin{equation}
    \int \textrm{d}V \phi(L) \qquad \phi(L)=\frac{A}{4\pi L^2}
\end{equation}
where $V$ is the volume of the detector, $\phi(L)$ the neutrino flux and $A$ the source activity. We can define the effective baseline $L_{eff}$ as the distance from the source at which a point-like detector of the same mass would observe the same event rate. $L_{eff}$ can be computed by solving
\begin{equation}
    \phi(L_{eff})M_D= \rho \int \textrm{d}V \phi(L)
\end{equation}
where $M_D$ is the detector mass and $\rho$ its density. This definition can be easily generalized to a finite-sized source, but the point-like case is sufficient for our purposes. For a spherical detector of radius $R$ and a source placed in the center, we would have $L_{eff}=R/\sqrt{3}$.  As a result, in this case if we increase the dimension of the detector the event rate would not increase as the mass (as $R^3$), but rather it would be proportional to $R$.  For a general source position, the event rate will not be directly proportional to $R$.  Since most backgrounds scale linearly with the mass of the detector, for very large detectors it would be optimal to not consider the whole fiducial volume, but rather only part of it, close to the source. 

We will consider 2 different detectors that could be used for this experiment: JUNO \cite{JUNO:2015zny,JUNO:2021vlw} and Jinping Neutrino Experiment (JNE) \cite{Jinping:2016iiq}, which are both liquid scintillator detectors.  JUNO is located in Jiangmen county \cite{noi} while JNE is in China's JinPing Underground Laboratory (CJPL). JUNO's construction is already completed, the detector was provisory filled with water which will be gradually replaced by LAB scintillator this year~\cite{JUNOtalk}. Regarding JNE, a 1-ton prototype is already operative, using a LAB-based scintillator \cite{Wang:2017ynm,Wu:2022oxo}.  In a few years, JNE will be upgraded to a 500 ton detector \cite{Wang:2024upf,SeminarWentai}. During Phase I, scheduled to run in 2026, the target material will be pure water, while in Phase II (2027-2036) it will be either LAB or a LiCl aqueous solution \cite{Guo:2017nnr,Shao:2022yjc}. All of the relevant information regarding the detectors considered, such as the mass, dimension and target material, can be found in Tab.~\ref{tab::Detector}.

 \begin{table}[htbp]
\centering
\caption{\protect\label{tab::Detector}  The mass, dimension and target material of each detector considered. For each row, the relevant reference is reported next to the name of the experiment.} 
\smallskip
\begin{tabular}{ccccc}
\hline
Detector& Mass (ton) & Geometry & R (m) &   Target Material \\
\hline
JUNO \cite{JUNO:2021vlw} & 20,000  & Sphere & 17.2  &   LAB  \\ \hline
JNE (prot.) \cite{Wu:2022oxo} & 1 & Sphere & 0.65 & LAB\\  \hline
JNE Phase I \cite{SeminarWentai} & 500 & Sphere &   5  &   Water \\  \hline
JNE Phase II \cite{SeminarWentai} & 500 & Sphere &   5  &   LAB or LiCl \\  \hline
\end{tabular}
\end{table}
We will first discuss the expected number of events, considering different source positions, then we will turn to the main backgrounds.

We considered two different configurations with the source placed outside or in the center of the detector. 
If the source is placed inside, its dimension could be relevant: we approximate it as a 20-cm sphere. This can be justified as follows.  At the BEST experiment, in order to achieve $10^{17}$ Bq activity, a 4 kg source was used. The density of Cr is about 7.2 gr/cm$^3$, which means that a 5.1 cm sphere would weigh approximately 4 kg. However, the 320 keV $\gamma$'s should be shielded. 
The total cross section for Compton scattering per electron can be computed using the Klein–Nishina formula and is around $3.5\times10^{-25}$ cm$^{2}$, while the total cross section for electron-neutrino scattering (again, computed for a single electron) is about $4.6\times 10^{-45}$ cm${}^2$, which means that one needs a suppression factor of at least $10^{20}$. The attenuation coefficient $\mu$ for 300 keV $\gamma$'s in lead is 4.57 cm$^{-1}$ \cite{AttenuationGamma}, which means that 15 cm of lead shielding should suppress $\gamma$ by a factor $\sim 10^{30}$, more than enough for our purposes.

Tab.~\ref{tab::Events0Th} reports the expected number of events for different detectors, source positions, and running time, assuming no low-energy threshold $T_L$, {\it i.e.} $T_L=0$. In Tab.~\ref{tab::Events200Th} the same quantities are reported, but assuming $T_{L}=200$ keV.
 \begin{table}[htbp]
\centering
\caption{\protect\label{tab::Events0Th}  Expected number of events for different detectors and source positions, assuming $T_{L}=0$. $D$ indicates the distance of the source from the center of the detector.} 
\smallskip
\begin{tabular}{crrrr}
\hline
Detector&  D (m) & Events/day & 10 days & 30 days \\
\hline
JUNO & 0 & 21,732.6 & 192,268.0 &458,533.4 \\ \hline
JUNO & 19 & 7,704.6 & 68,162.4 & 162,558.3  \\ \hline
JNE 1 ton & 0 & 533.0 & 4,715.0 & 11,244.7  \\ \hline
JNE 1 ton & 1 & 119.8 & 1059.7 & 2,527.1  \\ \hline
JNE 500 ton &0 & 6,244.8 & 55,247.7 & 131,758.5 \\ \hline
JNE 500 ton & 6 & 1,822.7 & 16,125.0 & 38,456.0 \\ \hline

\end{tabular}
\end{table}

 \begin{table}[htbp]
\centering
\caption{\protect\label{tab::Events200Th}  Same as Tab.~\ref{tab::Events0Th}, but with $T_{L}=200$ keV.} 
\smallskip
\begin{tabular}{crrrr}
\hline
Detector&  D (m) & Events/day & 10 days & 30 days \\
\hline
JUNO & 0 & 12,881.1 & 113,959.1 & 271,777.2\\ \hline
JUNO & 19 & 4,566.6 & 40,400.5 & 96,349.9\\ \hline
JNE 1 ton & 0 & 315.9 & 2,794.7 & 6,664.9 \\ \hline
JNE 1 ton & 1 & 71.0 & 628.1 & 1,497.86 \\ \hline
JNE 500 ton &0 & 3,701.4 & 32,745.9 & 78,094.5\\ \hline
JNE 500 ton & 6 & 1,080.3 & 9,557.5 & 22,793.3 \\ \hline

\end{tabular}
\end{table}

\subsection{Background}
Most of the backgrounds would be the same both at JNE and JUNO.  In the latter, however, one must also consider the cosmogenic muon background which is negligible at JNE due to its larger overburden \cite{JNE:2020bwn,JNE:2021cyb} (700 m rock at JUNO vs 2,400 m rock at JNE). For this reason, we will first discuss the backgrounds at JNE, then see what would change at JUNO.
At JNE, there are three possible target materials that could be used.  The 1 ton prototype, which is already operative, uses LAB as target material.  After the upgrade, in Phase I the target material will be purified water while in Phase II, it will be either LAB or LiCl dissolved in water. 

For both LAB and LiCl, the main source of background will be the radioactive decays of the unstable isotopes in the materials used, while this source of background is negligible for water. Indeed, the half-life of $^{15}$O, the only unstable isotope of oxygen that occurs naturally, is very low ($\sim 120$ seconds), which means that it can be safely ignored. In principle, it could still be produced inside the detector via interactions with cosmic rays.  The very large overburden, however, ensures that at CJPL the cosmic muons flux is extremely suppressed, and this source of background would be negligible as well. $^{3}$H could be, in principle, a source of background, since its half-life is a bit longer ($\sim 12$ years), however its Q-value is extremely low (18.6 keV), so it can be completely vetoed by a low-energy cut, which would not significantly affect our signal.

The situation would be more complicated if LiCl is dissolved in water, due to the presence of $^{36}$Cl. Its half-life is $3.0\times 10^5$ years, which means that it cannot be easily eliminated with a cooling period and its Q-value is 710 keV, which means its beta spectrum would completely overlap with our signal.  Its average abundance is $7\times 10^{-13}$, however depending on the provenience of the sample, this figure could vary, down to $10^{-16}$ \cite{Kazemi}. If such a solution is used, the proposed concentration would be 80 gr of LiCl in 100 gr of water \cite{Liang:2022lqn}. If the abundance of $^{36}$Cl is $7\times 10^{-13}$ ($10^{-16}$ ) the expected background rate would be $1.4\times 10^{10}$ ($2.0\times 10^6$) events/day for a 500 ton detector, far higher than our expected signal. In order to reduce this source of background to the level of 6,000 events/day (comparable with the signal rate, if the source is placed inside the detector), the abundance of $^{36}$Cl should be of the order of $3\times 10^{-19}$.

In the case of LAB, the main source of background would be $^{14}$C.  Its half-life is $5.7\times 10^3$ years and its natural abundance is $10^{-12}$.  However, it is possible to use carbon with a lower $^{14}$C abundance.  At Borexino, for example, the $^{14}$C abundance is $\sim 2\times 10^{-18}$ \cite{Borexino:1998eqi}. The event rate would still be very high.  If natural carbon is used, in a 500 ton detector, there would be $8.4\times 10^{12}$ events/day.  Assuming the same abundance as Borexino, this would be reduced to $1.7\times 10^{7}$ events/day, which would still be considerably higher than the signal rate. The good news is, however, that the Q-value of $^{14}$C is quite low, 156 keV, so a low-energy threshold could reduce this rate significantly. Please note that the exact value of the low-energy cut would depend on the ratio between the signal and background rates, as well as the energy resolution, and in general would be higher than 150 keV. 

Another possible way to veto this background would be to look for Cherenkov radiation, which can be distinguished from the scintillation light \cite{Li:2015phc}. Indeed, as an electron can emit Cherenkov light in LAB only if its kinetic energy is larger than 175 keV, and the Q-value of $^{14}$C is 156 keV, no electron emitted during this kind of decay can produce Cherenkov radiation.  We again stress that, it would anyway be difficult to detect Cherenkov light if the recoil energy is close to the threshold.  For this reason, we will assume $T_L=200$ keV if Cherenkov light is used to veto the $^{14}$C background.

\begin{figure}[ht]
\centering 
\includegraphics[width=0.8\linewidth]{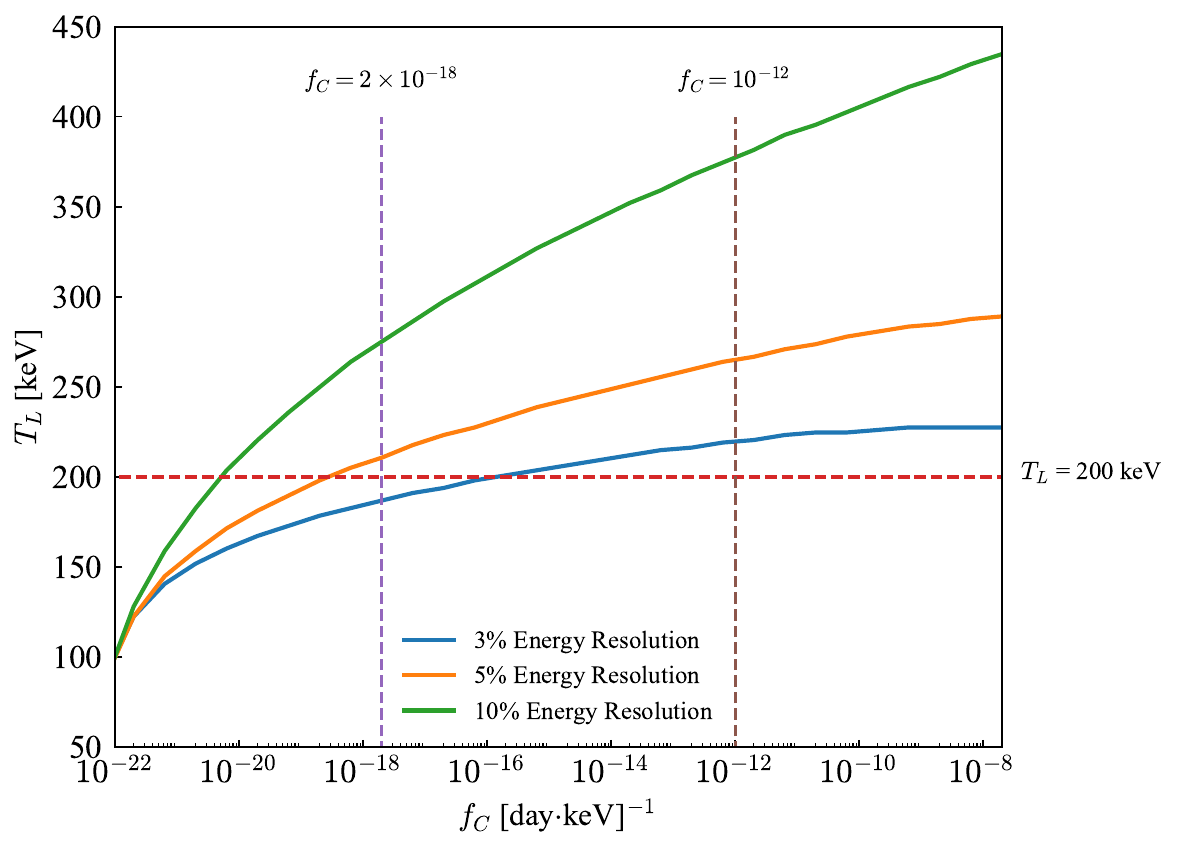}
\caption{\protect\label{fig:BackgroundInducedThreshold} Signal and background spectra. We used JNE 500 ton as the detector, with LAB as the target material and 5\% energy resolution. }
\end{figure}

Fig.~\ref{fig:BackgroundInducedThreshold} reports the value of the background-induced threshold (defined as the point where the event rate of the signal becomes larger than that of $^{14}$C decays), as a function of the abundance of $^{14}$C $f_C$.  For convenience, three lines have been added: two vertical ones, corresponding to the natural abundance of $^{14}$C, as well as the one used in Borexino, and a horizontal one, corresponding to 200 keV, {\it i.e.} the threshold at which Cherenkov light may be used to veto the $^{14}$C background.  The three curves correspond to energy resolutions of 3\%, 5\% (which is the target goal for JNE) and 10\%.  One can see that the background-induced threshold would be larger than 200 keV in all cases considered, except if Borexino's $^{14}$C abundance is used, and if the energy resolution is 3\%. This technique, however, could not be used for other sources of background, such as $^{36}$Cl.

Tab.~\ref{tab::MaterialBackgroun} summarizes the background rates from $^{36}$Cl and $^{14}$C with various abundances. For convenience, their half-life and Q-values are reported as well.
 \begin{table}[htbp]
\centering
\caption{\protect\label{tab::MaterialBackgroun}  Background rate from $^{36}$Cl and $^{14}$C, using various assumptions regarding their abundances $f_A$, and with a 500 ton effective mass.} 
\smallskip
\begin{tabular}{ccccc}
\hline
Isotope& Q-value & $t_{1/2}$ (yr) & $f_A$ & Rate (evts/day)  \\
\hline
$^{36}$Cl & 710 keV & 3.0$\times 10^5$ & $7\times 10^{-13}$ & $1.4\times 10^{10}$ \\ \hline
$^{36}$Cl & 710 keV & 3.0$\times 10^5$ & $1\times 10^{-16}$ & $2.0\times 10^6$\\ \hline
$^{36}$Cl & 710 keV & 3.0$\times 10^5$ & $3\times 10^{-19}$ & $6.0\times 10^3$\\ \hline
$^{14}$C & 156 keV & 5.7$\times 10^3$ & $1\times 10^{-12}$ & $8.4\times 10^{12}$\\ \hline
$^{14}$C & 156 keV & 5.7$\times 10^3$ & $2\times 10^{-18}$ & $1.7\times10^{7}$\\ \hline
\end{tabular}
\end{table}

Other sources of background would be subdominant, and usually lower (or comparable) to our signal.
\begin{itemize}
    \item {\bf Solar Neutrinos}. The main contribution would come from the $pp$ chain and from the 860 keV neutrino emitted in the $^7$Be decay, however they would both be subdominant \cite{Jinping:2016iiq}.  If the target mass is 500 ton, the estimated event rate for the former would be about 660-670 events/day, depending on the solar model (moreover, the maximum energy of $pp$ neutrinos is 420 keV, which means most of those events would be at very low energy), while the latter would be about 215-235 events/day.  Both would be significantly below our expected event rate
    \item {\bf Other radioactive background} from either radioactive contamination in the detector, the shielding, electronics, {\it{etc.}}.. or from isotopes created by interactions with cosmogenic muons. In JNE the event rate for these backgrounds is estimated to be, in total, of the order of 130 events/day \cite{Jinping:2016iiq} for a 500 ton detector, and can be safely ignored.
\end{itemize}

\begin{figure}[ht]
\centering 
\includegraphics[width=0.8\linewidth]{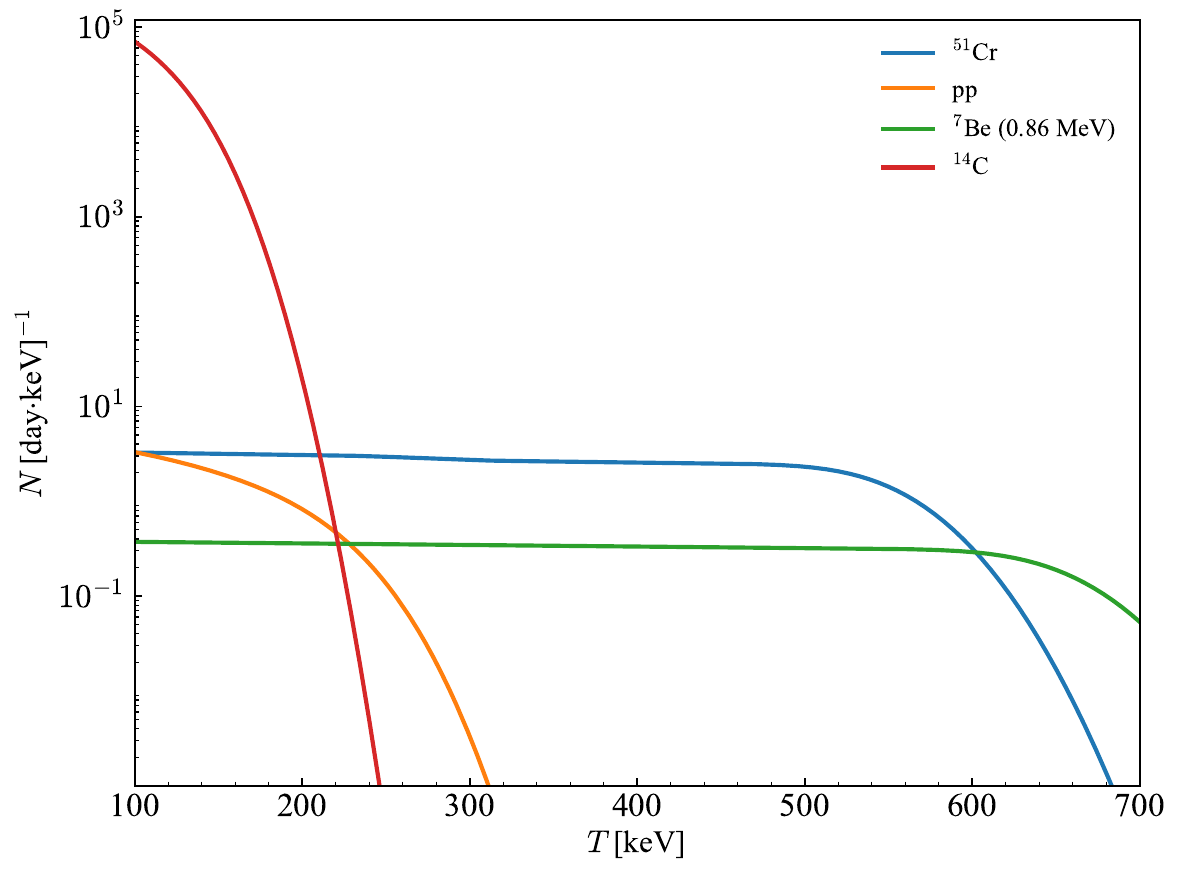}
\caption{\protect\label{fig:AllSpectra} Signal and background spectra. The 500 ton JNE detector was used, assuming LAB as target material 5\% energy resolution and $f_C=2\times 10^{-18}$. }
\end{figure}

Fig.~\ref{fig:AllSpectra} shows the expected event rate from $^{51}$Cr (assuming the source is placed in the center of the detector) as well as from the most relevant backgrounds. We used JNE 500 ton as a detector, LAB as target material and $\sigma_E/E=5\%/\sqrt{E/1\textrm{\ MeV}}$.

What would happen at JUNO? First of all, the solar neutrino background would increase much more than the signal because, as we discussed before, the former grows linearly with the mass, while the latter does not. For this reason, the expected number of events due to pp neutrinos is estimated to be around 27,600 events/day, while the $^{7}$B neutrino rate will be $\sim$10,000 events/day \cite{JUNO:2015zny}.

Moreover, due to the lower overburden, there is another source of background that might be relevant. Indeed the cosmogenic muon flux, despite being strongly suppressed, is considerably stronger at JUNO than at JNE. As a consequence, the rate with which their interactions with the LAB in the detector can create radioactive isotopes is much higher. In particular, $^{11}$C could be an issue: the end-point of its $\beta^+$ spectrum is 960 keV, which means it would cover the entire region of interest. While at JNE the $^{11}$C background rate would be negligible (less than 1 event/day \cite{Jinping:2016iiq}), at JUNO it would be significantly higher, namely $\sim$37,000 events/day \cite{JUNO:2015zny}. Both solar and $^{11}$C backgrounds could be reduced by considering only a fraction of the detector, which would increase the signal/background ratio. They would nonetheless be a significant issue for this experiment.

\section{Measurement of the Branching Ratio}
\label{sec::BR}
The branching ratio for $^{51}$Cr has already been measured with great precision.  However, a possible error in the BR has been cited as a candidate explanation for the Gallium anomaly.  In this section, we show that our proposal can, in some cases, provide an independent determination of the BR with sufficient precision to disfavor its role in the Gallium anomaly at the $2\sigma$ level.

\begin{figure}[ht]
\centering 
\includegraphics[width=0.8\linewidth]{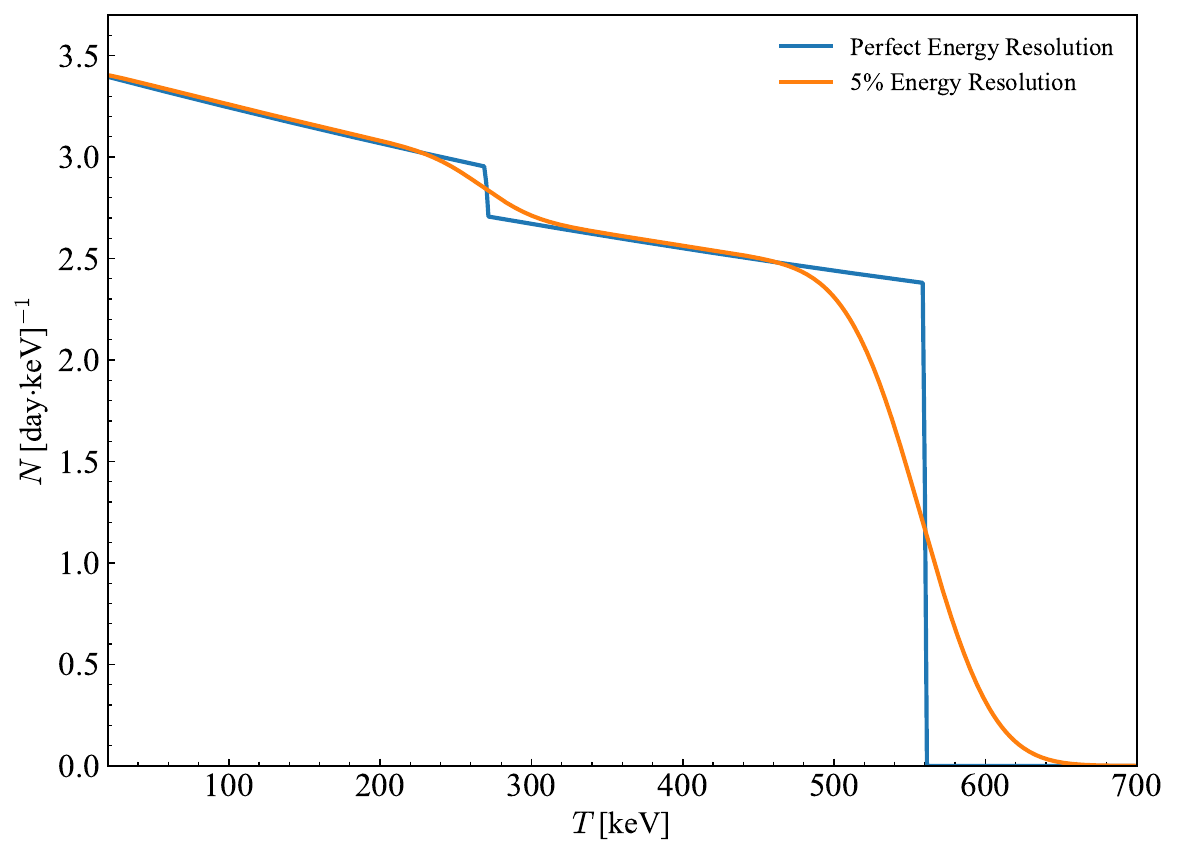}
\caption{\protect\label{fig:SpectrumRecoil} Recoil spectrum, with perfect and 5\% energy resolution.}
\end{figure}

The determination of the BR is more difficult and the requirements are considerably steeper. One approach is to consider the energy spectrum of the recoiled electron: since the cross section does not approach 0 when $T\rightarrow T_{max}$, one would have a corresponding ``step" in $T_{max,1}$, as can be seen in Fig.~\ref{fig:SpectrumRecoil}. 
Let us start by assuming perfect energy resolution. We can divide the spectrum into 2 bins, $N_{1}$ from $T_L$ to $T_{max,1}$, $N_2$ from $T_{max,1}$ to $T_{max,2}$\footnote{If the finite energy resolution is taken into account, the definition of the first bin can be left unchanged, but the second one should extend past $T_{max,2}$, with the exact limit to be determined taking into account the finite energy resolution and the background. However, since the number of events which would fall outside the bin is quite limited, when we will later discuss the finite energy resolution case, we will ignore this issue and assume that, regardless of the energy resolution, all the events with observed recoil energy higher than $T_{max,1}$ will still fall into $N_{2}$.}, with $N_{tot}=N_1+N_2$.
Both $N_2$ and $N_{tot}$ depend on $p_\mathrm{BR}$.  Setting $\tilde{p}_\mathrm{BR}=0.1$, assuming perfect energy resolution and  $T_{L}=200$ keV, we have
\begin{equation}
 N_{2}(p_\mathrm{BR})=(0.78-0.86(p_\mathrm{BR}-\tilde{p}_\mathrm{BR})) N_{tot}(\tilde{p}_\mathrm{BR}) \qquad  N_{tot}(p_\mathrm{BR})=(1-0.90(p_\mathrm{BR}-\tilde{p}_\mathrm{BR})) N_{tot}(\tilde{p}_\mathrm{BR}).
\end{equation}
Let $p_2=N_2/N_{tot}$ be the fraction of events that end up in the second bin and neglecting the next-to-leading order terms, one finds
\begin{equation}\label{eq::Defp2}
 p_\mathrm{BR}\simeq 0.1-6.15(p_{2}-0.78).
\end{equation}
Since, fixing $N_{tot}$, $N_2$ follows a Poisson distribution, we have
\begin{equation}\label{eq::SigTotSpectrum}
    \sigma_\mathrm{BR}\sim6.15\sqrt{\frac{p_2(1-p_2)}{N_{tot}}}\sim\frac{2.56}{\sqrt{N_{tot}}}.
\end{equation}
To measure the BR with 1\% precision, one must observe $\sim 65,000$ events, which would be challenging. It could be easily achieved in JUNO if the whole detector is used as a fiducial volume. However, in that case we would have issues with the cosmogenic background. It could be achieved in JNE as well: if the source is placed inside the detector and $T_L=200$ keV, in 30 days of running time we should see around 78k events, however as we will see the finite energy resolution will complicate the issue. 

Indeed, letting $q_1$ and $q_2$ be the fraction of events that migrate from bin 1 to bin 2 due to the energy resolution and vice versa, we have
\begin{equation}
    p_2\rightarrow p'_2=p_2(1-q_2)+(1-p_2)q_1.
\end{equation}
Eq.~(\ref{eq::SigTotSpectrum}) would now read
\begin{equation}
     \sigma_{\mathrm{BR}}\sim\frac{6.15}{1-q_1-q_2}\sqrt{\frac{p'_2(1-p'_2)}{N_{tot}}}.
\end{equation}
With a 5\% energy resolution, we would have $q_1\sim 0.14$ and $q_2\sim0.04$. In order to achieve $\sigma_{\mathrm{BR}}=1\%$, about 100,000 events are required.

On the other hand, measuring both the recoil energy $T$ and also the scattering angle $\theta$, one can reconstruct the neutrino energy using the relation
\begin{equation}\label{eq::DefEnu}
 E_\nu=\frac{m_e}{\textrm{Cos}(\theta)\sqrt{\frac{T+2m_e}{T}}-1}.
\end{equation}
Analogously to the $N_1$ and $N_2$ defined before, let  $n_{1(2)}$ the number of recoiled electrons generated by neutrinos with energy $E_{\nu,1(2)}$.  As before, $n_1+n_2=N_{tot}$. We can also write
\begin{equation}
    p_{\nu}=\frac{n_1}{N_{tot}}=\alpha p_{\mathrm{BR}} \qquad \alpha=\frac{\int_{T_L}^{T_{max,1}}\varphi_1(T)}{\int_{T_L}^{T_{max,2}}\varphi_T(T)}
\end{equation}
where $\varphi_i(T)$ is the differential event rate generated by neutrinos with energy $E_{\nu,i}$ and $\varphi_T(T)=p_{\mathrm{BR}}\varphi_1(T)+(1-p_{\mathrm{BR}})\varphi_2(T)$. $\varphi_i(T)$ is given by
\begin{equation}\label{def::VarPhi}
    \varphi_i(T)=\frac{d\sigma}{d T}(T,E_{\nu,i})\phi(L_{eff})\frac{M N_A}{m_D }n_{eff}
\end{equation}
where d$\sigma /$d$T$ is defined in Eq.~(\ref{eq::DefSigmaT}), M is the detector mass, $N_A$ is Avogadro's number, $m_D$ the molar mass of the target material and $n_{eff}$ is the effective duration of the experiment, considering the depletion of the source. 

If $T_L=200$ keV, one finds $\alpha\sim 0.19$, which implies that $p_\nu$ must be measured very precisely in order to achieve the desired accuracy in the measurement of $p_{\mathrm{BR}}$. It is worth noting that, even if $T_L=0$, one would still find $\alpha\sim 0.5$, due to the dependence of the total cross section on the neutrino energy.

Following the same procedure as before, let $q_1$ and $q_2$ be the fractions of events migrating from $n_1$ to $n_2$ and vice versa
\begin{equation} \label{eq::DefPprimo}
    n_1= p'_\nu N_{tot} \quad n_2=(1-p'_\nu)N_{tot} \quad p'_\nu=\alpha p_{\mathrm{BR}} (1-q_1-q_2)+q_2.
\end{equation}

$\sigma_{\mathrm{BR}}$ is now
\begin{equation}\label{eq::DefSigTheta}
    \sigma_{\mathrm{BR}}=\frac{1}{\alpha(1-q_1-q_2)}\sqrt{\frac{p'_\nu(1-p'_\nu)}{N_{tot}}}.
\end{equation}
One sees that, if $q_1=q_2=0$, this additional information on the neutrino energy greatly improves the precision. The reason is that $p_\nu$ is directly proportional to $p_{\mathrm{BR}}$ (unlike $p_2$, since there are constant terms in Eq.~(\ref{eq::Defp2}) ), which means that when $q_1=q_2=0$
\begin{equation}
    \sigma_{\mathrm{BR}}\sim \sqrt{\frac{p_{\mathrm{BR}}}{\alpha}}\frac{1}{\sqrt{N_{tot}}}=\frac{0.73}{\sqrt{N_{tot}}}.
\end{equation}
The computation of $q_1$ and $q_2$ depends on our choice of the criteria with which we will assign a certain neutrino energy to a given event. The most natural choice is, given a particular observed recoil energy $T_{obs}$ and an observed scattering angle $\theta_{obs}$, to compute the probability that the neutrino energy was $E_{\nu,i}$, namely $P(E_{\nu,i}|T_{obs},\theta_{obs})$. This can be computed using Bayes' theorem
\begin{equation}
    P(E_{\nu,i}|T_{obs},\theta_{obs})=\frac{P(T_{obs},\theta_{obs}|E_{\nu,i})P(E_{\nu,i})}{P(T_{obs},\theta_{obs})}.
\end{equation}
If $P(E_{\nu,1}|T_{obs},\theta_{obs})<P(E_{\nu,2}|T_{obs},\theta_{obs})$, all the events observed with $T=T_{obs}$, $\theta=\theta_{obs}$ will be assumed to be produced by a neutrino with $E=E_{\nu,2}$, or vice versa. Since $P(T_{obs},\theta_{obs})$ is a normalization constant that does not affect the calculations, we do not need to compute it. It should be stressed, however, that $P(E_{\nu,i})$ is not the probability of having a neutrino with energy $E_{\nu,i}$ (in other words: the BR), instead it is the probability that an electron-neutrino scattering event is caused by a neutrino with energy $E_{\nu.i}$, which means
\begin{equation}
   P(E_{\nu,i})=\left\{ \begin{array}{l l}
        \alpha p_{\mathrm{BR}} & i=1 \\
         1-\alpha p_{\mathrm{BR}} & i=2.
   \end{array} \right. 
\end{equation}
$P(T_{obs},\theta_{obs}|E_{\nu,i})$ is defined as
\begin{eqnarray}
    P(T_{obs},\theta_{obs}|E_{\nu,i})=&&\int_{T_L}^{T_{max,i}}\textrm{d}T \tilde{\varphi}_i(T)
    \frac{1}{\sqrt{2\pi \sigma_T(T)^2}}e^{-(T-T_{obs})^2/(2\sigma_T(T)^2)} 
    \times \nonumber\\
    && \frac{1}{\mathcal{R}(T,E_{\nu,i})\sqrt{2\pi \sigma_\theta(T)^2}}e^{-(\theta(T,E_{\nu,i})-\theta_{obs})^2/(2\sigma_\theta(T)^2)} \label{def::PtGivenEn}
\end{eqnarray}
where $\sigma_{T(\theta)}(T)$ is the precision of the measurement of $T$ $(\theta)$ which, in principle, could depend on $T$ and $\theta(T,E_{\nu,i})$ is the scattering angle corresponding to an electron recoil energy $T$ if the neutrino energy is $E_{\nu,i}$ which can be obtained by inverting Eq.~(\ref{eq::DefEnu}). $\mathcal{R}(T,E_{\nu,i})$ is a renormalization factor, due to the fact that $\theta_{obs}$ should be integrated from 0 to $\pi/2$ and is given by
\begin{equation}
    \mathcal{R}(T,E_{\nu,i})=\frac{1}{2} \left(\textrm{Erf}\left(\frac{\pi/2-\theta(T,E_{\nu,i})}{\sqrt{2} \sigma_\theta}\right)+\textrm{Erf}\left(\frac{\theta(T,E_{\nu,i})}{\sqrt{2} \sigma_\theta}\right)\right).
\end{equation}
Finally, $\tilde{\varphi}(T)$ is the event rate as defined in Eq.~(\ref{def::VarPhi}), but normalized to 1
\begin{equation}\label{def::TildeVarPhi}
    \tilde{\varphi}_i(T)=\frac{\varphi_i(T)}{\int_{TL}^{T_{max,i}}\textrm{d}T \varphi_i(T)}.
\end{equation}
Both the integrals in Eq.~(\ref{def::PtGivenEn}) and Eq.~(\ref{def::TildeVarPhi}) start from $T_L=200$ keV because we need to observe Cherenkov light to measure the scattering angle.  We recall that, regardless of the observed recoil, electrons with energy lower than 200 keV cannot produce enough Cherenkov light to be considered. 
The total number of events migrating from bin $i$ to bin $j$ ($i\neq j)$ is given by
\begin{equation}
    n_{i\rightarrow j}=N_{tot}\int_{\mathcal{X}_i}\textrm{d} T_{obs} \textrm{d} \theta_{obs} P(E_{\nu,i}|T_{obs},\theta_{obs}) P(T_{obs},\theta_{obs})=N_{tot}P(E_{\nu,i})\int_{\mathcal{X}_i}\textrm{d} T_{obs} \textrm{d} \theta_{obs} P(T_{obs},\theta_{obs}|E_{\nu,i}) 
\end{equation}
where $\mathcal{X}_i$ is defined as
\begin{equation}
\mathcal{X}_i=\{\forall (T_{obs},\theta_{obs}) | P(E_{\nu,i}|T_{obs},\theta_{obs})<P(E_{\nu,j}|T_{obs},\theta_{obs}), i\neq j \}
\end{equation}
From Eq.~(\ref{eq::DefPprimo}), one sees that $q_i$ is defined as $\frac{n_{i\rightarrow j}}{n_i}$, so there is an additional factor $N_{tot}/n_i=1/P(E_{\nu,i})$. We have
\begin{equation}
    q_i=\frac{N_{tot}}{n_i}P(E_{\nu,i})\int_{\mathcal{X}_i}\textrm{d} T_{obs} \textrm{d} \theta_{obs} P(T_{obs},\theta_{obs}|E_{\nu,i})=\int_{\mathcal{X}_i}\textrm{d} T_{obs} \textrm{d} \theta_{obs} P(T_{obs},\theta_{obs}|E_{\nu,i}) .
\end{equation}

Another notable feature is that, since $p_{\mathrm{BR}}=0.1$, it is very unlikely that $P(E_{\nu,1}|T_{obs},\theta_{obs})>P(E_{\nu,2}|T_{obs},\theta_{obs})$, which means that $q_2$ will be significantly suppressed. This is relevant because, following Eq.~(\ref{eq::DefPprimo}) and (\ref{eq::DefSigTheta}), if $q_2\ll 1$, $p'_\nu\simeq \alpha p_{\mathrm{BR}}(1-q_1)$, which means $\sigma_{\mathrm{BR}}\propto 1/\sqrt{\alpha}$ instead of $\sigma_{\mathrm{BR}}\propto 1/\alpha$.

\begin{figure}[ht]
\centering 
\includegraphics[width=0.8\linewidth]{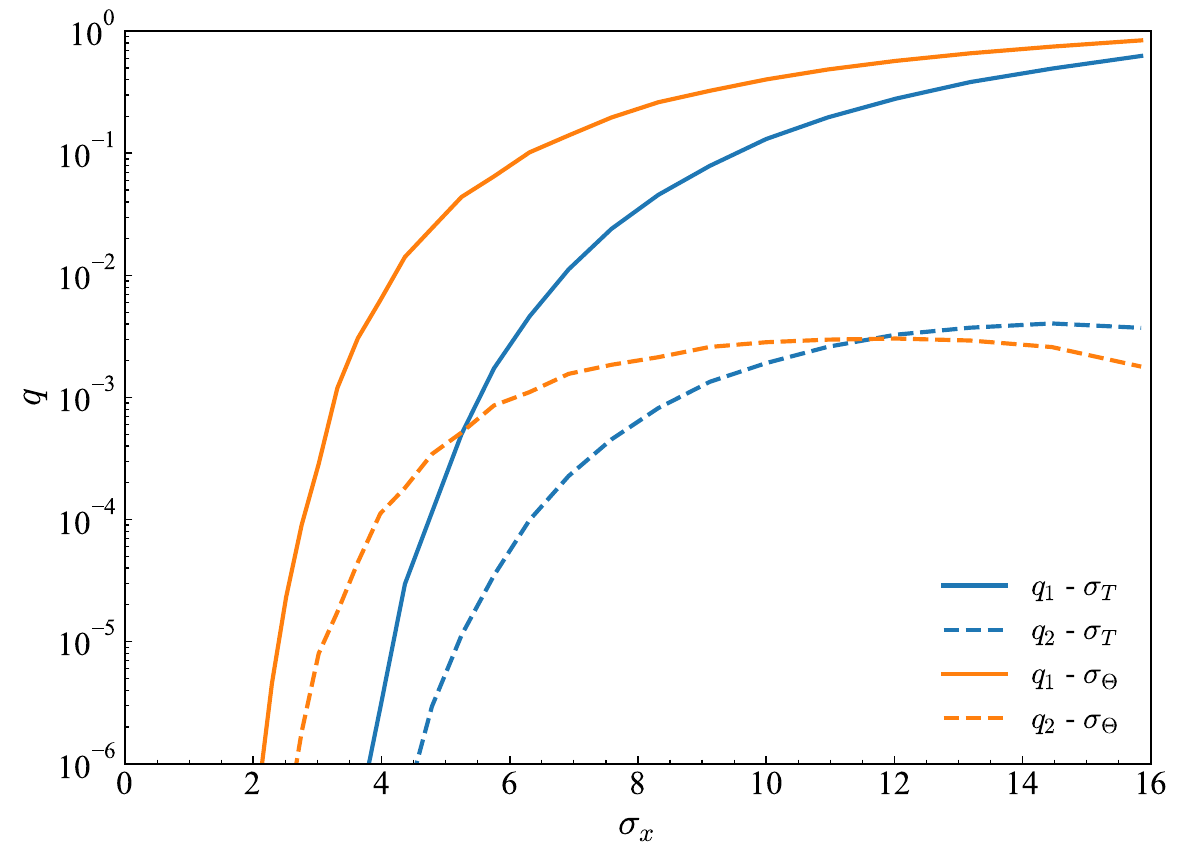}
\caption{\protect\label{fig:QvsSigma-Single} Value of $q_1$ and $q_2$ assuming either $\sigma_{T}$ or $\sigma_\theta$=0, as a function of the other. $\sigma_T$ is the value of $\sigma_T/T$, expressed in percentage and computed at 1 MeV (which means that 5 would correspond to $\sigma_T/T=5\%/\sqrt{T/1\textrm{\ MeV}}$, while $\sigma_\theta$ is expressed in degrees.}
\end{figure}

In Fig.~\ref{fig:QvsSigma-Single} we report the value of $q_1$, $q_2$ assuming either $\sigma_{T}$ or $\sigma_\theta$=0, as a function of the other. $\sigma_T$ is assumed to be proportional to $\sqrt{T}$, to reflect the statistical fluctuations in the number of photoelectrons. For simplicity, $\sigma_\theta$ is supposed to be independent of $T$. We see that $q_2$ is strongly suppressed and can be safely neglected, as expected. The main limitation will also come from the measurement of $\theta$ since at JNE, $\sigma_T/T$ is expected to be 5\% at 1 MeV \cite{Jinping:2016iiq}, which would make its contribution to $q_1$ negligible. In fact, in the range $200<T<560$ keV we would have $\sigma_E\sim 20-40$ keV. For comparison, if we compute the recoil energy corresponding to a given scattering angle for $E_{\nu,1}$ and $E_{\nu,2}$, their difference is less than 40 keV only when $\theta\sim60^\circ$.  However, this would correspond to a recoil energy of 53 and 93 keV, respectively, for which we would not be able to see Cherenkov light in any case. Finally, it is worth noting that we expect $q_2$ to decrease if the uncertainties are very large, as we can actually observe in Fig.~\ref{fig:QvsSigma-Single}. Indeed, the larger are the uncertainties, the less information we can obtain from $T_{obs}$ and $Q_{obs}$ on the neutrino energy. To clarify this point, we can consider the extremal case where, regardless of the true value of the recoil energy and the scattering angle, we would have a constant probability of observing any value of $T_{obs}$ and $\theta_{obs}$ in a finite range (in order to avoid the issues with normalization we would have if we simply send $\sigma_T\rightarrow\infty$). In this case, $P(T_{obs},\theta_{obs}|E_{\nu,i})$ would be a constant and more importantly, would not depend on $E_{\nu,i}$, which means that $P(E_{\nu,i}|T_{obs},\theta_{obs})=P(E_{\nu,i})$. In this extreme case, since we would not be able to obtain any information on the neutrino energy from the observables, following the procedure we described, we would be forced to assume that every event would be caused by a neutrino with energy $E_{\nu,2}$, simply because those events are more likely, and there would be no events migrating from $n_2$ to $n_1$, which means that $q_2=0$.

\begin{figure}[ht]
\centering 
\includegraphics[width=0.8\linewidth]{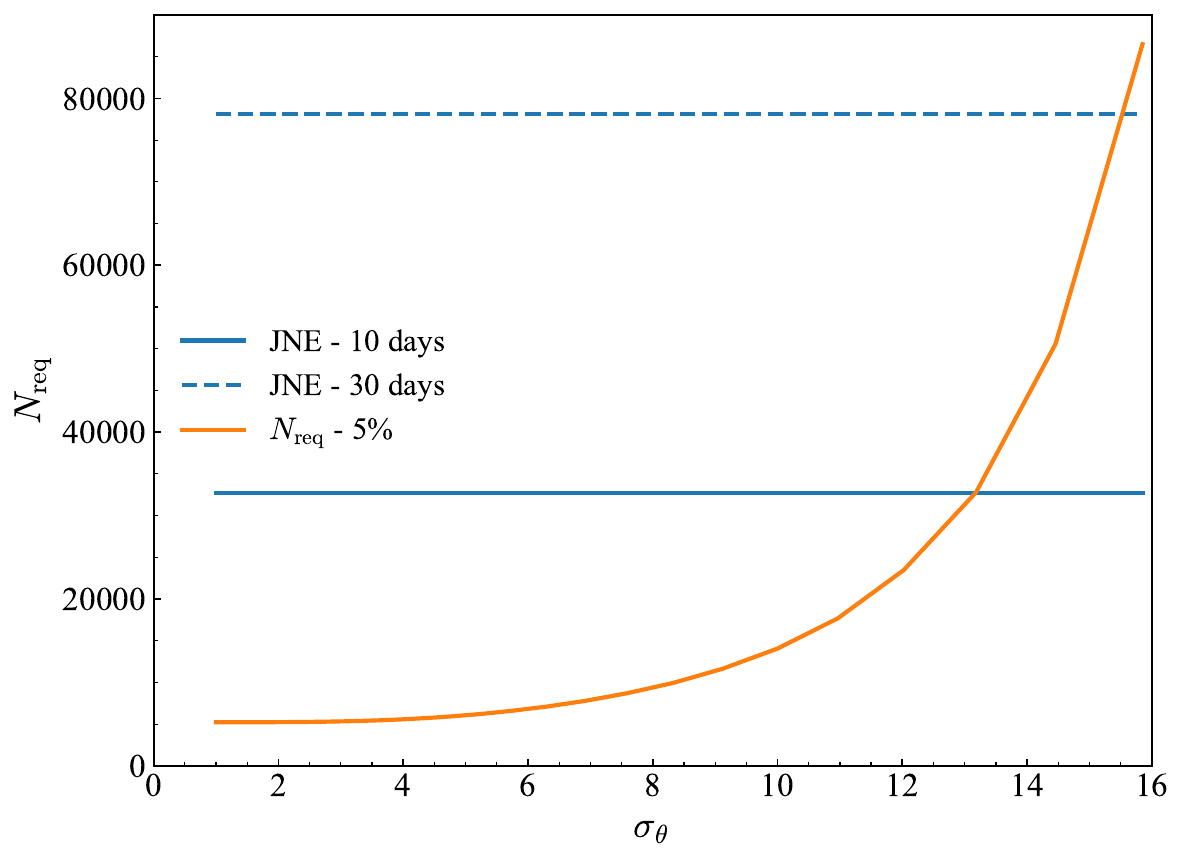}
\caption{\protect\label{fig:NreqVsSigma-Tot} Total number of events required to achieve $\sigma_{\mathrm{BR}}=1\%$, computed according to Eq.~(\ref{eq::DefSigTheta})}
\end{figure}

In Fig.~\ref{fig:NreqVsSigma-Tot}, we report the total number of events required to achieve $\sigma_{\mathrm{BR}}=1\%$, assuming the energy resolution of 5\% and as a function of $\sigma_\theta$. We can see that by measuring the scattering angle, it would be possible, at least in principle, to measure $p_{\mathrm{BR}}$ with sufficient accuracy to exclude at 2$\sigma$ an error in the estimation of the source activity as the origin of the anomaly. However, it should be noted that in \cite{Luo:2022xrd} the authors computed the angular resolution assuming a recoil energy of 2 MeV and 5.6\% energy resolution, finding a value of $\sigma_\theta\sim 46^\circ$, which is considerably larger than would be needed to achieve this result with 10 days of data taking at significantly lower energies (we would need $\sigma_\theta\leq 13^\circ$ for recoil energies between 200 and 560 keV). This could be a serious problem, that is not easily overcome, because such an accuracy would be limited not only by the PTM coverage, but also by the amount of Cherenkov light produced, which is very little at those energies.
Other approaches to measure the scattering angle would also be very challenging. If a TPC is used, the target material should be in gaseous form because the tracks in liquid TPC's would be too short to measure the scattering angle reliably. However, with HP TPC, the density of the target material would be considerably lower, and the effective baseline would appreciably suppress the expected number of events.  Moreover, it would not be possible to place the source inside the detector, further decreasing the flux.

\section{Baseline Dependence}
\label{sec::Sterile}
Using liquid scintillators, it would also be possible to study the baseline dependence of the anomaly, if present. Indeed, if we observe that the missing flux has some sort of dependence on the baseline, we would be able to assume that the cause of the anomaly is related to the propagation of neutrinos. The most popular mechanism for such an effect is the mixing with one or more families of sterile neutrinos. However, as we mentioned earlier, such a result would be in tension with several other sterile neutrino experiments, which probed the same region of parameter space we will be able to investigate and failed to find any evidence of sterile neutrinos, so additional mechanisms would be needed to explain the tension \cite{Brdar:2023cms}. 

\begin{figure}[ht]
\centering 
\includegraphics[width=0.8\linewidth]{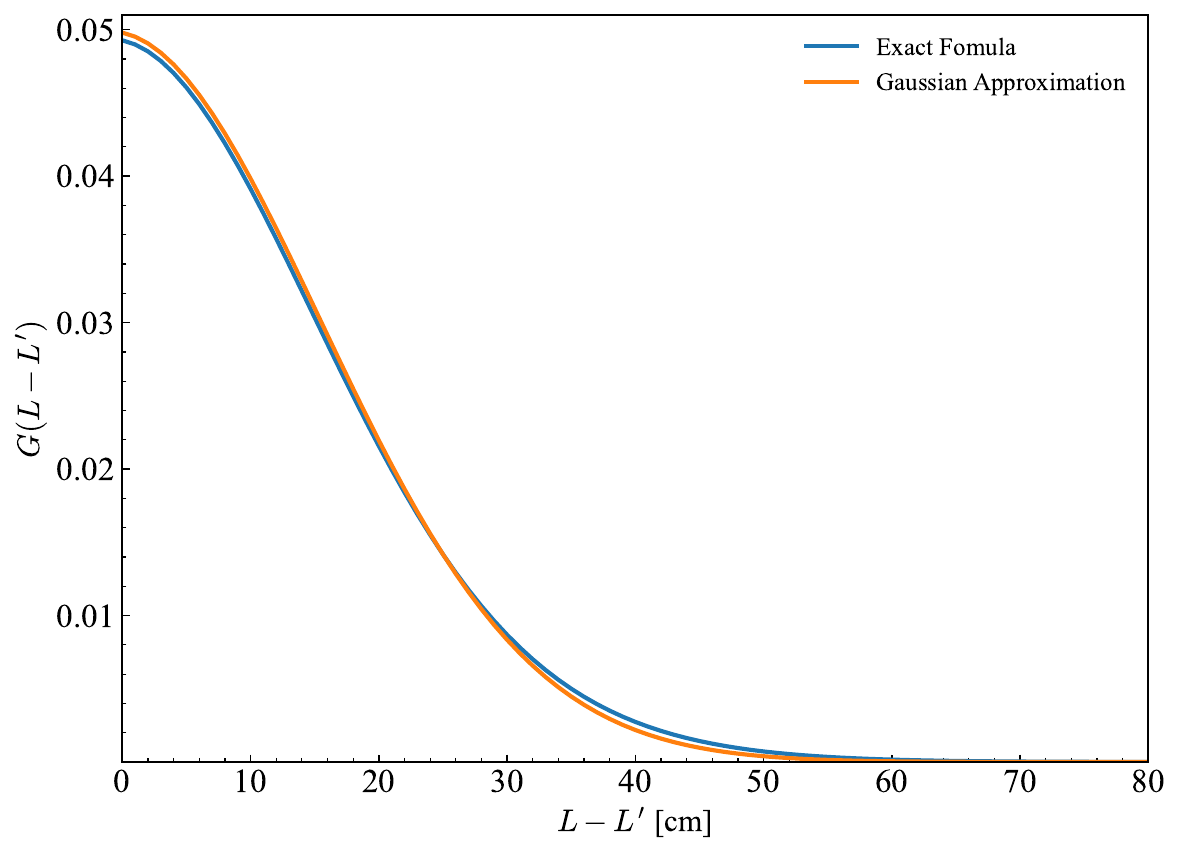}
\caption{\protect\label{fig:Gauss} $G(L-L')$ computed using the exact formula reported in Eq.~(\ref{def::GLL}) and a Gaussian approximation with $\sigma_L=16.4$ cm, computed using Eq.~(\ref{def::sigmaL}).}
\end{figure}

To search for sterile neutrinos, we can exploit the finite dimension of the detector and the close proximity of the source. We can divide the detector into $n$ bins, and the $i-$th bin will contain all the events observed at distances between $L_{i-1}$ and $L_{i}$ from the source. Let $N_{0,i}$ be the expected number of events if there are no sterile neutrinos, and let $N_{\nu,i}(\Delta m^2,\theta)$ denote the same quantity, computed assuming the existence of a sterile neutrino with mass square difference $\Delta m^2$ and mixing angle $\theta$. We have
\begin{equation}
    N_{0,i}=N_{tot}\frac{\int_{L_{i-1}}^{L_{i}} \textrm{d}L \int \textrm{d}L'\frac{dV}{dL'}\phi(L')G(L-L')}{\int_{L_0}^{L_{n}} \textrm{d}L  \frac{dV}{dL}\phi(L)}
\end{equation}
\begin{equation}
    N_{\nu,i}(\Delta m^2,\theta)=N_{tot}\frac{\int_{L_{i-1}}^{L_{i}} \textrm{d}L \int_{L_0}^{L_n} \textrm{d}L' P(L',\Delta m^2,\theta) \frac{dV}{dL'}\phi(L')G(L-L')}{\int_{L_0}^{L_{n}} \textrm{d}L \frac{dV}{dL}\phi(L)}.
\end{equation}
$dV/dL$ depends on the position of the source.  Let $R$ be the detector radius and, if the source is outside, let $D$ be the distance between the source and the center of the detector.  We then have
\begin{eqnarray}
   \textrm{ source in the center} &\rightarrow& \frac{dV}{dL}=4\pi L^2    \\
   \textrm{ source at distance D} &\rightarrow& \frac{dV}{dL}=2\pi L^2\left(1-\frac{D^2+L^2-R^2}{2D L} \right).   
\end{eqnarray}
If the source is inside and has dimension $d=20$ cm,\footnote{In principle, the finite dimension of the source could also contribute to the uncertainty on the baseline, we will not take this into account, however.} $L_0=d$ and $L_n=R$.  If it is outside, we have $L_0=D-R$, $L_n=D+R$. $G(L-L')$ represents the spatial resolution of the detector.  This depends on the recoil energy of the electron.  We would have $\tilde{\sigma}_L(T)=10 \textrm{cm}/\sqrt{T/\textrm{1 MeV}}$ \cite{Smirnov:2020bcr}, which means that, in principle, $G(L-L')$ should be defined as
\begin{equation}\label{def::GLL}
    G(L-L')=\frac{\int_{T_L}^{T_{max,2}}\textrm{d}T \varphi_T(T)\frac{1}{\sqrt{2\pi \tilde{\sigma}^2_L(T)}}e^{-(L-L')^2/2\tilde{\sigma}^2_L(T)}}{\int_{T_L}^{T_{max,2}}\textrm{d}T \varphi_T(T)}.
\end{equation}
For our purposes, however, we can assume that $G(L-L')$ is a standard Gaussian function, with $\sigma_L$ defined as
\begin{equation}\label{def::sigmaL}
    \frac{1}{\sigma_L^2}=\frac{\int_{T_L}^{T_{max,2}}\textrm{d}T \frac{\varphi_T(T)}{\tilde{\sigma}^2_L(T)}}{\int_{T_L}^{T_{max,2}}\textrm{d}T \varphi_T(T)}\quad \Rightarrow \quad \sigma_L=16.4\textrm{ cm}.
\end{equation}
In Fig.~\ref{fig:Gauss}, we report $G(L-L')$, both calculated according to Eq.~(\ref{def::GLL}) and using a Gaussian approximation with $\sigma_L=16.4$ cm.

$P(L',\Delta m^2,\theta)$ is defined as
\begin{equation}
    P(L,\Delta m^2,\theta)=1-\textrm{Sin}^2(2\theta)\left(0.019\textrm{Sin}^2\left(\frac{1.27\Delta m^2 L}{E_{\nu,1}}\right)+0.981\textrm{Sin}^2\left(\frac{1.27\Delta m^2 L}{E_{\nu,2}}\right) \right).
\end{equation}
Here $\Delta m^2$ should be in eV$^2$, $L$ in m and $E_{\nu,i}$ in MeV. Using the Asimov data set \cite{Cowan:2010js}, $\Delta \chi^2$ is defined as
\begin{equation}
    \Delta\chi^2(\Delta m^2,\theta)=\sum_{i=1}^n\frac{(N_{\nu,i}(\Delta m^2,\theta)-N_{0,i})^2}{N_{0,i}}
\end{equation}
and the 2-$\sigma$'s regions are defined as the region of the parameter space where $\Delta\chi^2(\Delta m^2,\theta)<6.18$.

\begin{figure}[ht]
\centering 
\includegraphics[width=0.8\linewidth]{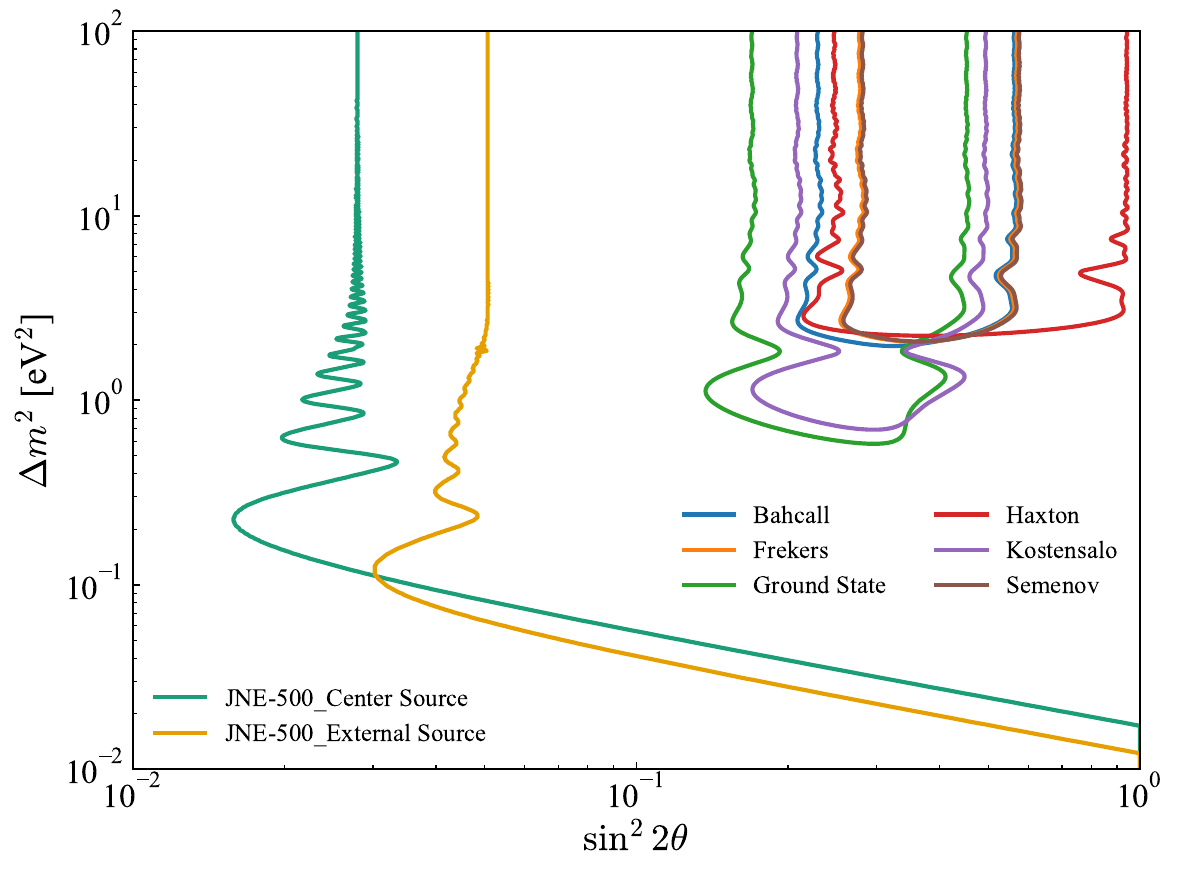}
\caption{\protect\label{fig:SterileNeutrinos} Exclusion contours for sterile neutrino (2$\sigma$'s), using 500 ton JNE as detector and 10 days of exposure. For a comparison, the allowed region for the Gallium Anomaly as reported in \cite{Giunti:2022btk} is also shown.}
\end{figure}

We considered only JNE-500, due to the lower background.  The dimensions of the detector can be found in Tab.~\ref{tab::Detector}.  When the source was outside, we used $D=6$ m. We also assumed $T_L=200$ keV and a running time of 10 days, using the values of $N_{tot}$ from Tab.~\ref{tab::Events200Th}. The resulting exclusion contours can be found in Fig.~\ref{fig:SterileNeutrinos}.

\section{Conclusions}
\label{sec::Conclusions}

We discussed the possibility of testing the Gallium Anomaly, observing the $\nu_e$ via electron-neutrino scattering in liquid scintillators instead of using Ga detectors. In this way, if the anomaly is still present, we would know that it is not due to any effect related to the detection process; conversely, if it disappears, we would know where its origin lies. 

In order to determine the neutrino flux with sufficient precision, the detector should be at least of the order of tons and have a large overburden. The detector should also be compact, because the effective baseline scales like its dimension $R$: as a consequence, the expected number of events grows as $R$, however the background usually is proportional to the mass, {\it i.e.}  to $R^3$.  As a result, if the detector is very large, it could be convenient to consider only part of the fiducial volume, close to the source. We considered two candidates: JUNO (20k tons, 700 m rock overburden) and JNE, which is located in CJPL (2,400 m rock overburden); a 1-ton prototype of JNE is already operative, a 500 ton detector will be completed in the next years. Assuming a 200 keV low-energy threshold, the event rate would span form 71 events/day (JNE-1 ton, if the source is outside the detector) up to 12.9k events/day (JUNO, if the source is in the center). This means that even at JNE-1 ton it would be possible to observe the anomaly with 5 $\sigma$'s confidence level with 10 days of exposure, taking into account the source depletion. 

We found that at JUNO cosmogenic muon background could be a concern: the expected event rate for $^{11}$C events is of the order of 37k events/day, significantly larger than the signal rate. The impact of $^{11}$C background could be mitigated by considering only part of the detector, vetoing the rest, while it is negligible at JNE, due to the larger overbuden. The main source of background in most cases would be unstable isotopes in the target material. For example, if LAB is used, the $^{14}$C atoms inside would provide a significant background rate. However, the end-point of the beta spectrum for this decay is 156 keV and the electrons emitted would not be able to emit Chernekov light: this can be used to reject this background completely, which would impose a {\it de facto} low-energy threshold, rejecting also all the $^{51}$Cr events with recoil energy lower than 200 keV. Another possible way to determine the low-energy threshold, without using Cherenkov light, would be to look at the point where the $^{14}$C background is larger than the signal, which would depend on the background and signal event rates, as well as on the energy resolution. However, due to the large statistical fluctuations present at low recoil energies, this kind of threshold would be higher than 200 keV in most of the cases.

Even if the origin of the anomaly is not in the Gallium detectors, we could still collect important information. By looking at the shape of the spectrum and/or by measuring the scattering angle as well, it is possible to measure the BR of $^{51}$Cr, which would be useful to exclude any systematic errors in the estimation of the source activity. The requirements for such a measurement, however, would be much steeper: if we use only the spectral information, even assuming perfect energy resolution we would need 65k events to determine the BR with an accuracy of 1\%, which would exclude this possible explanation for the anomaly only at 2 $\sigma$'s. The finite energy resolution would further increase the the number of events needed up to 100k, considerably more of what we could expect to see at JNE-500 ton.  If the source is placed in the center of the detector and with 30 days of exposure we would observe 78k events. Measuring the scattering angle would decrease the number of events needed, however the precision required would be considerably higher than what we can reasonably achieve: $\sigma_\theta$ should be lower than 14$^\circ$ (16$^\circ$) degrees in the relevant energy range (200-560 keV) if we want to determine the BR with 1\% of accuracy using 10 (30) days of exposure at JNE-500. By comparison, the expected precision when the recoil energy is 2 MeV is around 46$^\circ$. For this kind of measurement, then, we would need a stronger source, or a considerably better angular resolution. 

\section*{Acknowledgement}
EC grateful to Wentai Luo, Marco Grassi and Carlo Giunti for the useful discussions and suggestions. JT was supported in part by National Natural Science Foundation of China under Grant Nos. 12347105 and Fundamental Research Funds for the Central Universities (23xkjc017) in Sun Yat-sen University. EC is supported by Gansu Provicial Grant for Senior Foreigner Experts 24RCKA006. JT is grateful to Southern Center for Nuclear-Science Theory (SCNT) at Institute of Modern Physics in Chinese Academy of Sciences for hospitality.  JE is supported by the Ministry of Education, Science, Culture and Sport of the Republic of Armenia under the Remote Laboratory Program, grant number 24RL-1C047. 


\end{document}